\documentclass[twocolumn,showpacs,showkeys,superscriptaddress]{revtex4}
\usepackage{graphicx}
\usepackage{amsmath}
\usepackage{amsfonts}
\usepackage{amssymb}
\usepackage{mathrsfs}

\def\openone{\leavevmode\hbox{\small1\kern-3.8pt\normalsize1}}
\def\N{\leavevmode\hbox{ Z \kern-8 pt\normalsize{Z}}}
\def\openone{\leavevmode\hbox{\small1\kern-3.8pt\normalsize1}}
\def\openJ{\leavevmode\hbox{J \kern-9.5pt\normalsize J}}
\def\openS{\leavevmode\hbox{ S \kern-9.3pt\normalsize S}}
\newcommand{\bb}{\begin{equation}}
\newcommand{\ee}{\end{equation}}
\newcommand{\eqb}{\begin{eqnarray}}
\newcommand{\eqf}{\end{eqnarray}}

\usepackage{color}

\begin{document}

\title{Electromagnetic redshift in anisotropic cosmologies}

\author{Sergio A. Hojman}
\email{sergio.hojman@uai.cl}
\affiliation{UAI Physics Center, Universidad Adolfo Ib\'a\~nez, Santiago, Chile.}
\affiliation{Departamento de Ciencias, Facultad de Artes Liberales,
Universidad Adolfo Ib\'a\~nez, Santiago, Chile.}
\affiliation{Centro de Investigaci\'on en Matem\'aticas, A.C., Unidad M\'erida; Yuc, M\'exico.}
\affiliation{Departamento de F\'{\i}sica, Facultad de Ciencias, Universidad de Chile,
Santiago, Chile.}
\affiliation{Centro de Recursos Educativos Avanzados,
CREA, Santiago, Chile.}
\author{Felipe A. Asenjo}
\email{felipe.asenjo@uai.cl}
\affiliation{UAI Physics Center, Universidad Adolfo Ib\'a\~nez, Santiago, Chile.}
\affiliation{Facultad de Ingenier\'{\i}a y Ciencias,
Universidad Adolfo Ib\'a\~nez, Santiago, Chile.}

\begin{abstract}
The redshift of light is calculated for an anisotropic cosmological spacetime. Two different approaches are considered. In the first one, electromagnetic waves are modeled using the geometrical optics (high--frequency) approximation. This approach considers light rays following null geodesics, being equivalent to the motion followed by pointlike spinless massless particles. It is shown that the redshift for this case depends, in general, on the direction of propagation,  and is dispersive (wavelength dependent) for light emitted from different points of an extended object. In the second approach electromagnetic waves are studied using the exact form of Maxwell equations, finding that redshift has dependence on the direction of propagation as well as on the wave polarization. The electromagnetic waves are dispersive and depend on the anisotropic temporal evolution. In this last case, redshifts may become dispersive   depending on the relative direction between the light ray propagation vector and the anisotropy axes.  The relation of these results with a possible violation of the Equivalence Principle are discussed. In general, both results are set in the context of recent astrophysical redshift observations for anisotropic cosmologies, and new ways of determining redshifts are suggested.
\end{abstract}


\maketitle

\section{Introduction}

Constants of motion are fundamental tools for solving differential equations.
With those, physical sensible information can be extracted easily from the studied models. In curved spacetimes, some constants can be found by using Killing vectors, which, for example, are essential for understanding the redshift suffered by light in cosmological scenarios. A Killing vector $\xi_{\mu}$ is defined by the equation \cite{misner,wald}
\begin{equation}\label{killingeq}
\nabla_\mu \xi_{\nu}+\nabla_\nu \xi_{\mu}=0\, ,
\end{equation}
 where $\nabla_\mu$ stands for covariant differentiation. Finding a Killing vector makes it possible to determine conserved quantities along the geodesics of observers. Thus, the knowledge of a Killing vector allows us to define conserved quantities that may be measured by those observers.

For instance, for a given metric, consider a momentum wavevector $K^\mu$ which is parallel transported along geodesics. Thereby,  the first integrals $C$,  determined by the  Killing vectors, are  given by
\begin{equation}\label{constanKilli}
C=\xi_\mu K^\mu\, .
\end{equation}

In this work, we use the Killing vectors associated to Bianchi I anisotropic cosmologies to study the propagation of light in those settings, and at the same time,  to determine the redshift of light. We will study how the anisotropic structure of spacetime introduces new effects in the redshift, and how they can be used as an experimental tool to determine any kind of anisotropy encoded in the current cosmological observations.

For time--dependent spacetimes there are no timelike Killing vectors, and thus energy is not conserved. Therefore, only spacelike Killing vectors can be used to define constants of motion associated to spacelike features of any electromagnetic wave.
In  here, we consider the Bianchi I cosmological model \cite{ryan} in cartesian coordinates, representing a general anisotropic  expanding Universe described by the metric
\begin{equation}\label{binchibiachi}
g_{\mu\nu}= {\mbox {diag}} \ [-1,a^2(t),b^2 (t),c^2 (t)]\, ,
\end{equation}
where, in general, every spatial direction has different time-dependent scale-factors $a(t)$, $b(t)$ and $c(t)$, denoting the anisotropic expansion of the Universe. The isotropic flat Friedmann-Robertson-Walker (FRW) cosmology is a particular case of Bianchi I spacetimes, for which $a(t)=b(t)=c(t)$.

For Bianchi I cosmology, there are three Killing vectors satisfying Eq.~\eqref{killingeq}. These are
\begin{eqnarray}\label{killingVectors}
{\xi}_\mu^{1}&=&(0,a^2,0,0)\, ,\nonumber\\
{\xi}_\mu^{2}&=&(0,0,b^2,0)\, ,\nonumber\\
{\xi}_\mu^{3}&=&(0,0,0,c^2)\, ,
\end{eqnarray}
which reduce to ${\xi}_\mu^{1}=(0,a^2,0,0)$, ${\xi}_\mu^{2}=(0,0,a^2,0)$, and ${\xi}_\mu^{3}=(0,0,0,a^2)$ for the isotropic FRW cosmology. The importance of the Killing vectors \eqref{killingVectors}, and their main physical difference with FRW cosmologies, is that they establish preferred directions on space (differently to the FRW case where every direction is equivalent). Those directions are determined by the cosmological model under consideration, and they define  principal axis on spacetime. Therefore, any physical measurable quantity can be studied using projections onto those axis.

In the subsequent sections, we  study the effect of those preferred directions on the space in the redshift of light. We show that redshift is highly dependent on the direction of propagation of the electromagnetic waves, giving rise to different redshifts as  light propagates in the anisotropic medium. In order to study the light dynamics thoroughly, we will consider the redshift produced by light following null geodesics \cite{misner} and by electromagnetic (EM) waves which do not evolve along geodesics \cite{hojman1,hojman2}. Light following null geodesics are EM waves which satisfy the geometrical optics approximation,  such that  its propagation is described as a light ray, i.e., spinless and massless pointlike particles \cite{misner}.

 On the contrary, if the geometrical optics or eikonal approximation conditions are not met, and the EM wave is studied by exactly solving Maxwell equations (without using the eikonal approximation), then it can be shown that EM waves do not, in general, follow geodesics \cite{hojman1,hojman2},  presenting also a coupling between its polarization and the curvature of the spacetime. This  implies that if general EM solutions of Maxwell equations are considered, then the redshift must be corrected due to this non--geodesic behavior.

 We will show how both dynamical settings give rise to different redshifts, and how they can be used to determine the properties of light propagating on different cosmological spacetimes.

\section{Anisotropic redshift for light propagating along null geodesics}
\label{sectionnull}

The EM vector potential $A_\mu (x^\nu)$ may be described in terms of real fields, its vectorial amplitude $\Sigma_\mu(x^\nu)$ and its phase $S(x^\nu)$ by $A_\mu  = \Sigma_\mu e^{iS}$, as usual. Consider an EM wave in the geometrical optics limit \cite{misner}. This approximation is performed by studying EM waves in the high--frequency limit, where all the variations of the amplitude of the wave are neglected in comparison with its frequency (this assertion is precisely stated in the following Section). Thus, light does not behave as a wave under this approximation.
In this case, light is modelled as spinless and massless point--like particles moving along rays
which follow null geodesics,  with the dispersion relation
\begin{equation}\label{geo1}
K_\mu K^\mu=0\, ,
\end{equation}
where $K_\mu=\partial_\mu S$ is the four--wavevector of the EM wave, and it is defined through the derivative of the (real function) phase $S(x^\mu)$ of an EM wave. This is equivalent to the assumption of a lightlike line--element $ds^2=g_{\mu\nu}dx^\mu dx^\nu=0$.
It is straightforward to show that \eqref{geo1} implies null geodesics propagation \cite{misner,carroll}
\begin{equation}\label{geo2}
K^\nu \nabla_\nu K^\mu=0\, .
\end{equation}

On the other hand, three constants $C^i$ can be constructed by using the three Killing vectors \eqref{killingVectors}
\begin{equation}\label{threeconstanKilli}
C^i=\xi_\mu^i K^\mu\, .
\end{equation}
These are constants along the null geodesic of the light ray, as it can be shown
\begin{equation}
K^\alpha \nabla_\alpha C^i=K^\alpha K^\mu \nabla_\alpha\xi_\mu^i +\xi_\mu^i K^\alpha \nabla_\alpha K^\mu=0\, ,
\end{equation}
where the first term vanishes identically due to Eq.~\eqref{killingeq}, while the second one is zero because of \eqref{geo2}. In the case of a light ray, the  constants correspond to the three independent components of the three--dimensional wavevector
\begin{eqnarray}\label{geo3}
C^1&=&\xi_\mu^1 K^\mu=g^{\mu\nu}\xi_\mu^1 K_\nu=K_x\, ,\nonumber\\
C^2&=&\xi_\mu^2 K^\mu=g^{\mu\nu}\xi_\mu^2 K_\nu=K_y\, ,\nonumber\\
C^3&=&\xi_\mu^3 K^\mu=g^{\mu\nu}\xi_\mu^3 K_\nu=K_z\, .
\end{eqnarray}
Thus, the spatial derivatives of the phase of the light wave are constant. The phase is linear in the three spatial directions  defined  by the anisotropy.

Now, let us consider an observer at rest with four--velocity $u^\mu=(-1,0,0,0)$, such that this observer measures a frequency given by $-u^\mu K_\mu\equiv \omega$. In this way, and considering the constants \eqref{geo3},
 the null vector $K_\mu$ can be explicitly written as \cite{wald}
\begin{equation}
K_\mu=\omega \ u_\mu+\frac{K_x}{a^2}\xi_\mu^1+\frac{K_y}{b^2}\xi_\mu^2+\frac{K_z}{c^2}\xi_\mu^3\, ,
\label{Kmu}
\end{equation}
as $u^\mu \xi_\mu^i=0$.  Contracting Eq. \eqref{Kmu} by $K^\mu$, and using Eq.~\eqref{geo1}, we get \cite{carroll}
\begin{equation}\label{conseBianLight}
-\omega\left(K^\mu u_\mu\right) =\frac{K_x}{a^2}\left( K^\mu \xi_\mu^1\right)+\frac{K_y}{b^2}\left( K^\mu \xi_\mu^2\right)+\frac{K_z}{c^2}\left( K^\mu \xi_\mu^3\right)\, .
\end{equation}
Thus, we can readily obtain the dispersion relation \eqref{geo1} that governs the propagation of light in the geometrical optics approximation
\begin{equation}\label{conseBianLight}
\omega=\left(\frac{K_x^2}{a^2}+\frac{K_y^2}{b^2}+\frac{K_z^2}{c^2}\right)^{1/2}\, ,
\end{equation}
from it is deduced that the observed frequency $\omega$ depends on time.
Hence, by using \eqref{geo3} and \eqref{conseBianLight} we can deduce the redshift of light.

In general, the cosmological redshift $z$ is defined as
\begin{equation}\label{redshix000}
z=\frac{\omega(t_e)}{\omega(t_o)}-1\, .
\end{equation}
where two freely falling observers   measuring different frequencies are needed, one of which observes light when is emitted at time $t_e$ and the other one which observes light at a later time $t_o$.
Now, as an example, consider a light ray propagating in the $x$-direction in such a way that $K_y=0=K_z$, and thus with frequency $\omega=K_x/a$. The wavevector $K_x$ is a constant along the null geodesic. From \eqref{redshix000}, the redshift for a light ray propagating in the $x$-direction is given by
\begin{equation}\label{redshix}
z=\frac{a(t_o)}{a(t_e)}-1\, .
\end{equation}
This result may seem to be straightforwardly expected, but it is not. In order to fully understand the complexity of this result, we need to explore the possibility of a light ray propagating in a null geodesic along the $y$-direction with frequency $\omega=K_y/b$ (where now $K_x=0=K_z$). In this direction of propagation, the redshift is now
\begin{equation}\label{redshiy}
z=\frac{b(t_o)}{b(t_e)}-1\, ,
\end{equation}
which is different, in general, from redshift \eqref{redshix}, as $a\neq b$. Evidently, a light ray propagating  along the $z$-direction, will also have a different redshift given by
\begin{equation}\label{redshiz}
z=\frac{c(t_o)}{c(t_e)}-1\, .
\end{equation}
The three redshifts reported above \eqref{redshix}, \eqref{redshiy} and \eqref{redshiz} are different, in general. This implies that redshift depends on the direction of propagation of light rays in an anisotropic cosmology. Any difference in the values of cosmological redshifts (for waves propagating in different directions) may be an indication of a preferred direction in the Universe.
 This is completely different from what ocurs in FRW cosmologies. When $a=b=c$, the three previous redshifts coincide for an isotropic Universe \cite{misner,wald,carroll,ryden}.

In this way, in an anisotropic cosmological model, light rays moving along null geodesics propagate differently in different directions, the redshift now depends strongly on direction and special care must be taken when interpretation of measurements are advanced. This can be simply exemplified for the case of a model of a Universe with small anisotropy $a=c$ and $b=a(1+\epsilon)$ with $\epsilon=\epsilon(t)\ll 1$ \cite{oron}. We have chosen the anisotropy in the $y$-direction, but it can be, of course, in any direction, in general.  This case is of special relevance, as we will discuss in the last section, there is  observational evidence that our Universe is almost isotropic, but with a   window for a small anisotropy yet undetected by current experimental capabilities \cite{ferreras}. So, to keep it simple, let us assume that $a(t_o)\approx 1$ and that the anisotropy is only on the past of the Universe, i.e., the current observed anisotropy $\epsilon(t_o)\approx 0$.
This is the case of the vacuum--dominated  Kasner solutions that isotropize the Universe for large times, even if it was originally anisotropic \cite{oron}.

First, consider light rays moving on the principal axes of the spacetime. For light rays in the $x$ or $z$-directions ($K=K_x$ or $K=K_{z}$), using the dispersion relation \eqref{conseBianLight}, we obtain
\begin{equation}\label{dispersionGeoGeo0}
\omega\approx\frac{K}{a}\, ,\qquad z=\frac{1}{a(t_e)}-1\, ,
\end{equation}
and those light rays suffer only the isotropic FRW--like redshift. The small anisotropy of the Universe does not affect the dynamics of light rays moving in such directions.  On the other hand, if the light ray is moving along the $y$-direction (with $K=K_y$), then its redshift
\begin{equation}\label{redshiG2}
\omega\approx\frac{K}{b}\, ,\qquad z\approx\frac{1}{a(t_e)}\left[1-\epsilon(t_e)\right]-1\, ,
\end{equation}
contains information  about the anisotropy of the Universe. Clearly, light rays do not propagate in the same way in all directions, and different redshifts are a consequence of that.

However, a more important consequence occurs for the case of light rays propagating in directions which are different from those of the principal axes. In general, let us define for a light ray with a given wavector, the parameter $K=(K_x^2+K_y^2+K_z^2)^{1/2}$. For  the small anisotropy case, the dispersion relation \eqref{conseBianLight} becomes
\begin{equation}\label{dispersionGeoGeo}
\omega\approx\frac{K}{a}\left(1-\frac{K_y^2}{K^2}\epsilon\right)\, ,
\end{equation}
and the redshift between the emitted and observed  frequency is
\begin{equation}\label{redshiG}
z\approx\frac{1}{a(t_e)}\left(1-\frac{K_y^2}{K^2}\epsilon(t_e)\right)-1\, ,
\end{equation}
meaning that the redshift is direction dependent, i.e., anisotropic. Only light rays propagating in a plane orthogonal to the anisotropic axis show an isotropic FRW redshift.  Furthermore, the anisotropic behavior of the redshift \eqref{redshiG} does not appear on an isotropic FRW background (with $\epsilon=0$), where no preferred direction of propagation exists, implying a direction independent redshift.

Accordingly, in the anisotropic model, if the redshifts $z_A$ and $z_B$ are determined for light emitted from two different points $A$ and $B$ in the same extended astronomical object, then $z_A\neq z_B$ in general. This occurs as the direction of propagation (wavevectors) of the radiated light could be different, being able to produce dispersion for the respective redshifts.

In this way, any measurement of different redshifts for light released by the same extended astronomical object (after considering the redshifts associated to rotation of the object), would be an indicator of anisotropy of the Universe.

\section{Anisotropic redshift for light described by electromagnetic waves which do not propagate along null geodesics}
\label{sectionnonnull}

As it was mentioned in the preceding section, the null geodesic behavior of light is obtained by using the geometrical optics approximation (shown below). However, when Maxwell equations are studied beyond that approximation, it can be proved that the null geodesics behavior of light does not, in general, hold \cite{hojman1,hojman2}.

Maxwell equations $\nabla_\alpha F^{\alpha\beta}=0$ can be written in terms of the four-vector potential $A_\mu$ as \cite{misner}
\begin{equation}\label{ecMcompl}
\frac{1}{\sqrt{-g}}\partial_\alpha\left[\sqrt{-g}g^{\alpha\mu}g^{\beta\nu} \left(\partial_\mu A_\nu-\partial_\nu A_\mu\right) \right]=0\, ,
\end{equation}
where $g$ is the metric determinant. We study an EM wave described by the four--potential $A_\mu=\Sigma_\mu \exp(iS)$ \cite{misner,carroll}, where $\Sigma_\mu$ is the amplitude and $S$ is the phase (both real), and with wavevector defined as $K_\mu=\partial_\mu S$. Then, from Eq.~\eqref{ecMcompl} we get  two evolution equations for the wavevector and for the amplitude
\begin{eqnarray}\label{ecMcompl1}
&&\left(K_\mu K^\mu\right)\Sigma^\beta-\left(K_\mu \Sigma^\mu\right) K^\beta=\nonumber\\
&&\quad\qquad\frac{1}{\sqrt{-g}}\partial_\alpha\left[\sqrt{-g}g^{\alpha\mu}g^{\beta\nu} \left(\partial_\mu \Sigma_\nu-\partial_\nu \Sigma_\mu\right) \right]\, ,
\end{eqnarray}
and
\begin{eqnarray}\label{ecMcompl2}
&&\frac{1}{\sqrt{-g}}\partial_\alpha\left[\sqrt{-g}\left(K^\alpha \Sigma^\beta-K^\beta \Sigma^\alpha\right)\right]+\nonumber\\
&&\qquad\qquad g^{\beta\nu}K^\mu\left(\partial_\mu \Sigma_\nu-\partial_\nu \Sigma_\mu\right)=0\, .
\end{eqnarray}
It can be shown that $K_\mu K^\mu=0$ is not, in general, an exact solution to the above equations \cite{hojman1,hojman2} (unless the geometrical optics approximation is used).

For the subject under study, let us consider the anisotropic spacetime \eqref{binchibiachi}. Also, let us assume transversal propagation with $K_\mu \Sigma^\mu=0$,  with variables depending on time only, and $u^\mu \Sigma_\mu=-\Sigma_0=0$. These conditions are consistent with the Lorenz gauge $\nabla_\mu A^\mu=\left[-\partial_0(\sqrt{-g}\ \Sigma_0)/\sqrt{-g}+i \Sigma^\mu K_\mu\right]\exp(iS)=0$.
 Thus, Eqs.~\eqref{ecMcompl1} and \eqref{ecMcompl2} simplify to
\begin{eqnarray}\label{ecMcompl3}
\left(K_\mu K^\mu\right)\Sigma^\beta=-\frac{1}{\sqrt{-g}}\partial_0\left[\sqrt{-g}g^{\beta\nu} \partial_0 \Sigma_\nu \right]\, ,
\end{eqnarray}
and
\begin{eqnarray}\label{ecMcompl4}
\frac{1}{\sqrt{-g}}\partial_0\left(\sqrt{-g} \ \omega \Sigma^\beta\right)+ \omega g^{\beta\nu}\partial_0 \Sigma_\nu=0\, .
\end{eqnarray}
The equations \eqref{ecMcompl3} and \eqref{ecMcompl4}, that describe the propagation of a EM wave in an anisotropic scenario, are now coupled. Notice that the amplitude depends on the frequency of the wave. This is a typical characteristic of a dispersive medium, such as an anisotropic cosmological spacetime. The geometrical optics approximation is reached when the right--hand side of Eq.~\eqref{ecMcompl3} is neglected, i.e., when the amplitude variations are negligible compared to the frequency of the wave, giving $K_\mu K^\mu=0$.

In particular, from Eq.~\eqref{ecMcompl3} we can find that the EM wave solutions of Maxwell equations have a dispersion relation of the form \cite{hojman1}
\begin{equation}\label{geoNon1}
K_\mu K^\mu= -\frac{\Sigma_\beta}{\sqrt{-g}\, \Sigma^\alpha \Sigma_\alpha}\partial_0\left[\sqrt{-g}g^{\beta\nu} \partial_0 \Sigma_\nu \right] \equiv\chi\, ,
\end{equation}
where, in our case, $\chi=\chi(t)$ is a time-dependent function, which does not vanish in general. The sign of $\chi$ depends on the explicit form of the anisotropic metric and on the polarization of the EM wave \cite{hojman1}; different EM wave polarizations give rise to different $\chi$. Furthermore, from Eq.~\eqref{ecMcompl4}, we can obtain the conservation equation
\begin{eqnarray}\label{ecMcompl5}
\partial_0\left(\sqrt{-g} \ \omega \Sigma^\beta\Sigma_\beta\right)=0\, ,
\end{eqnarray}
from where we can obtain the exact solution for the amplitude of the EM wave
\begin{eqnarray}\label{ecMcompl6}
\Sigma^\beta\Sigma_\beta=\frac{\mbox{constant}}{\sqrt{-g} \ \omega}\, .
\end{eqnarray}

These EM waves, which are  exact solutions to Maxwell equations, described by Eqs.~\eqref{ecMcompl3}, \eqref{ecMcompl4}, \eqref{geoNon1} and \eqref{ecMcompl6}, contain the information of the wave nature of light, i.e, its extended structure on spacetime. Therefore, the EM waves
do not follow geodesics (not even the null ones) in general. This can be proved by taking the derivative of \eqref{geoNon1} to obtain
\begin{equation}\label{geoNon2}
K^\alpha\nabla_\alpha K_\mu=\frac{1}{2}\partial_\mu\chi\, .
\end{equation}
This is a natural feature for an extended object. One can wonder whether an EM wave which does not follow null geodesics violates the Equivalence Principle (EP). The key to understand what is happening is to recognize that the EP is valid for pointlike objects only. Structured physical objects (either massive or massless) have physical extension (such as a wave) and/or internal degrees of freedom (such as spin) that must be taken into account. In those cases, there are several geodesic curves passing through the object and it experiences tidal forces. Thus, the EP is no longer applicable to extended structured objects.
When the geometrical optics approximation is invoked to solve Maxwell equations, light is modelled as a pointlike spinless and                         massless physical entity (light rays), travelling along null geodesics according to the EP. However, if Maxwell equation are solved beyond that limit, the extended size and internal structure (polarization) of the EM wave modifies its dynamics (as compared to that of a pointlike object). As a result, light described by an EM wave does not follow geodesics, in general.

Anyway one can find conserved quantities along the EM wave propagation.
In fact, the three quantities \eqref{threeconstanKilli} are still constant in this model along the curve whose tangent is the four--wavevector of the EM wave. This can be easily seen by calculating
\begin{equation}
K^\alpha \nabla_\alpha C^i=\frac{1}{2}\xi_\mu^i\partial^\mu \chi\equiv 0\ .
\end{equation}
The last term vanishes identically because $\chi$ is time-dependent only, and the time components of the Killing vectors vanish. Thereby, the three components of the three--dimensional wavevector \eqref{geo3} are always constants of motion.

In this way we can follow a similar procedure than previous section to define the wavevector.  The final result (which differs from that for a light ray) is
the dispersion relation \eqref{geoNon1} for an exact EM wave
\begin{equation}\label{conseBianLightNon}
-\omega^2+\frac{K_x^2}{a^2}+\frac{K_y^2}{b^2}+\frac{K_z^2}{c^2}=\chi\, .
\end{equation}
depending on the EM wave polarization through $\chi$. From this result, a general redshift  can be readily calculated
\begin{equation}
z=\sqrt{\frac{\left.\frac{K_x^2}{a^2}+\frac{K_y^2}{b^2}+\frac{K_z^2}{c^2}-\chi\right|_{t_e}}{\left.\frac{K_x^2}{a^2}+\frac{K_y^2}{b^2}+\frac{K_z^2}{c^2}-\chi\right|_{t_o}}}-1\, .
\end{equation}
These results show that, in general, the redshift depends on the dispersive properties of the EM wave (through its wavevectors or wavelenghts) and its polarization (through $\chi$).

In order to put this result in terms of an
explicit expression for wave propagation, let us consider the case of small anisotropy.
When the anisotropy has the form  $a=c$, $b=a(1+\epsilon)$, and $\epsilon\ll 1$, in the $y$-direction, then waves propagate differently in each direction \cite{hojman1}. Let us calculate the different redshifts for EM waves with polarizations aligned along the principal axes of the metric.
First, let us work out the case of polarization in the $x$-direction. Other directions for EM wave polarizations can be dealt with in an analogous fashion. Thus, consider an amplitude with the form $\Sigma_\mu=(0,\Sigma_x,0,0)$, and the wavevector $K_\mu=\omega \ u^\mu+{K_y}\xi_\mu^2/{b^2}+{K_z}\xi_\mu^3/{c^2}$, such that $K_\mu\Sigma^\mu=0$. Thus, the EM wave propagates on the $y-z$ plane. As the anisotropy is small, we consider a small departure $\eta_x=\eta_x(t)$ from the FRW EM frequency
\begin{equation}
\omega_x\approx\frac{\sqrt{K_y^2+K_z^2}}{a}\left(1+\eta_x\right)\, ,
\end{equation}
where $\eta_x\ll 1$. In this way, the amplitude $\Sigma_x$ can be obtained by solving Eqs.~\eqref{ecMcompl4} or \eqref{ecMcompl6}, to yield
\begin{equation}\label{amplitudEMwavex}
\Sigma_x\approx\frac{\mbox{ constant}}{\left(K_y^2+K_z^2\right)^{1/4}}\left(1-\frac{\epsilon+\eta_x}{2}\right)\, .
\end{equation}
The behavior of $\eta_x$ can be obtained from the dispersion relation \eqref{geoNon1} or \eqref{conseBianLightNon}. That relation gives rise to the  equation
\begin{equation}\label{etaxeq}
\frac{d^2 \eta_x}{d\tau^2}+4\left(K_y^2+K_z^2\right)\eta_x+\frac{d^2 \epsilon}{d\tau^2}+4 K_y^2\epsilon=0\, ,
\end{equation}
where we have introduced the FRW time
\begin{equation}
\tau=\int_0^t \frac{dt}{a(t)}\, .
\end{equation}
Several important cases can now be studied. First, the geometrical optics limit can be recovered from Eq.~\eqref{etaxeq} when  variations  of amplitude are neglected compared with the scales of the EM wave, i.e., ${d_\tau^2 \eta_x}/\eta_x\ll K_y^2+K_z^2$, and ${d_\tau^2 \epsilon}/\epsilon \ll K_y^2$. In this case, the solution of \eqref{etaxeq} is simply $\eta_x=- K_y^2\epsilon/( K_y^2+K_z^2)$, which is the result  \eqref{dispersionGeoGeo} for light in the geometrical optics limit. Notice that this fact occurs for any polarization.

Secondly, if the EM wave propagates in the $y$--direction only, then $K_z=0$ and Eq.~\eqref{etaxeq} has the solution $\eta_x=-\epsilon$. In this case, the wave propagates along null geodesics, with constant amplitude, and frequency and redshift coinciding with those presented in Eqs.~\eqref{redshiG2}.

Finally, if the EM waves propagate in a general form in the $y-z$ plane, then the solution of Eq.~\eqref{etaxeq} is
\begin{widetext}
\begin{eqnarray}\label{solutionetaX}
\eta_x(t)&=&\frac{\cos\left(2 \tau\sqrt{K_y^2+K_z^2}\right)}{2\sqrt{K_y^2+K_z^2}}\int_0^\tau \left[\frac{\partial^2\epsilon(v)}{\partial v^2}+4K_y^2\epsilon(v)\right]\sin\left(2v\sqrt{K_y^2+K_z^2}\right) dv\nonumber\\
&&-\frac{\sin\left(2 \tau\sqrt{K_y^2+K_z^2}\right)}{2\sqrt{K_y^2+K_z^2}}\int_0^\tau \left[\frac{\partial^2\epsilon(v)}{\partial v^2}+4K_y^2\epsilon(v)\right]\cos\left(2 v\sqrt{K_y^2+K_z^2}\right) dv\, ,
\end{eqnarray}
\end{widetext}
and therefore the redshift $z_x$ for an EM wave polarized in the $x$--direction can be readily calculated to be
\begin{equation}\label{redshiftEMwavexpolar}
z_x\approx \frac{1}{a(t_e)}\left[1+\eta_x(t_e)\right]-1\, ,
\end{equation}
where we have assumed that $a(t_o)\approx 1$ and that the current observed anisotropy vanishes $\epsilon(t_o)= 0$  [therefore $\eta_x(t \rightarrow t_o)\rightarrow 0$]. EM wave redshifts are more general than those for light rays, and they are dispersive, as $K_y\neq K_z$
in general. Besides, notice that $\eta$ depends on the temporal variation of $\epsilon$, through second-order time derivatives. Thus, this redshift contains information of the local temporal evolution of the anisotropic structure of the cosmological spacetime.

We can perform a similar analysis for an EM wave polarized in the $y$--direction, which propagates in the $x-z$ plane, in general. In this case, it is straightforward to show that the wave amplitude has the form
\begin{equation}\label{amplitudEMwavey}
\Sigma_y\approx \frac{\mbox{ constant}}{\left(K_x^2+K_z^2\right)^{1/4}}\left(1+\frac{\epsilon-\eta_y}{2}\right)\, ,
\end{equation}
where $\eta_y=\eta_y(t)$ is the small correction to the frequency of the $y$--polarized EM wave
\begin{equation}
\omega_y\approx\frac{\sqrt{K_x^2+K_z^2}}{a}\left(1+\eta_y\right)\, ,
\end{equation}
due to its non--geodesic behavior and the anisotropic spacetime (with $\eta_y=\eta_y(t)\ll 1$).
From the dispersion relation \eqref{geoNon1} or \eqref{conseBianLightNon} we can find the equation for the evolution of the small correction
\begin{equation}\label{etaxeq2}
\frac{d^2 \eta_y}{d\tau^2}+4\left(K_x^2+K_z^2\right)\eta_y-\frac{d^2 \epsilon}{d\tau^2}=0\, .
\end{equation}
The geometrical optics limit implies that $\eta_y\approx 0$, which coincides with the null geodesics propagation described by Eqs.~\eqref{dispersionGeoGeo0}. However, if the EM wave is studied beyond that limit, the behavior of $\eta_y$ is completely different. The solution of Eq.~\eqref{etaxeq2} is
\begin{widetext}
\begin{eqnarray}\label{solutionetaY}
\eta_y(t)&=&-\frac{\cos\left(2 \tau\sqrt{K_x^2+K_z^2}\right)}{2\sqrt{K_x^2+K_z^2}}\int_0^\tau \frac{\partial^2\epsilon(v)}{\partial v^2}\sin\left(2v\sqrt{K_x^2+K_z^2}\right) dv\nonumber\\
&&+\frac{\sin\left(2 \tau\sqrt{K_x^2+K_z^2}\right)}{2\sqrt{K_x^2+K_z^2}}\int_0^\tau \frac{\partial^2\epsilon(v)}{\partial v^2}\cos\left(2 v\sqrt{K_x^2+K_z^2}\right) dv\, .
\end{eqnarray}
\end{widetext}
Thus, the redshift  $z_y$ associated to an EM wave polarized in the $y$--direction becomes
\begin{equation}\label{redshiftEMwaveypolar}
z_y\approx \frac{1}{a(t_e)}\left[1+\eta_y(t_e)\right]-1\, .
\end{equation}
Notice that, again, this redshift is dispersive and more general that those for light rays. Also, as $\eta_x\neq \eta_y$, the redshifts \eqref{redshiftEMwavexpolar} and \eqref{redshiftEMwaveypolar} are different in general, and thereby, for EM waves, redshifts depend
on the wave polarizations.

This effect can also been obtained for an EM wave polarized in the $z$--direction
 propagating in the $x-y$ plane, such that $K_\mu \Sigma^\mu=0$. The EM wave has the frequency
\begin{equation}
\omega_z\approx\frac{\sqrt{K_x^2+K_y^2}}{a}\left(1+\eta_z\right)\, ,
\end{equation}
where $\eta_z=\eta_z(t)\ll 1$ is the correction due to the anisotropy to be determined. Its amplitude, through Eq.~\eqref{ecMcompl4}, can be shown to be
\begin{equation}\label{amplitudEMwavez}
\Sigma_z\approx \frac{\mbox{constant}}{\left(K_x^2+K_y^2\right)^{1/4}}\left(1-\frac{\epsilon+\eta_z}{2}\right)\, ,
\end{equation}
and using the dispersion relation  \eqref{conseBianLightNon}, we can obtain the  equation
\begin{equation}\label{etaxeq2}
\frac{d^2 \eta_z}{d\tau^2}+4\left(K_x^2+K_y^2\right)\eta_z+\frac{d^2 \epsilon}{d\tau^2}+4 K_y^2\epsilon=0\, .
\end{equation}
Anew, the geometrical optics limit can be recovered when ${d_\tau^2 \eta_z}/\eta_z\ll K_x^2+K_y^2$, and ${d_\tau^2 \epsilon}/\epsilon \ll K_y^2$, giving $\eta_z=- K_y^2\epsilon/( K_x^2+K_y^2)$. Furthermore, when the EM wave propagates in the $y$--direction only (with $K_x=0$), then $\eta_z=-\epsilon$, recovering the results of Sec.~\ref{sectionnull} for null geodesics propagation.
For a  general propagation in the $x-y$ plane, the solution of Eq.~\eqref{etaxeq2} is
\begin{widetext}
\begin{eqnarray}\label{solutionetaZ}
\eta_z(t)&=&\frac{\cos\left(2 \tau\sqrt{K_x^2+K_y^2}\right)}{2\sqrt{K_x^2+K_y^2}}\int_0^\tau \left[\frac{\partial^2\epsilon(v)}{\partial v^2}+4K_y^2\epsilon(v)\right]\sin\left(2v\sqrt{K_x^2+K_y^2}\right) dv\nonumber\\
&&-\frac{\sin\left(2 \tau\sqrt{K_x^2+K_y^2}\right)}{2\sqrt{K_x^2+K_y^2}}\int_0^\tau \left[\frac{\partial^2\epsilon(v)}{\partial v^2}+4K_y^2\epsilon(v)\right]\cos\left(2 v\sqrt{K_x^2+K_y^2}\right) dv\, ,
\end{eqnarray}
\end{widetext}
and the redshift $z_z$ that an EM wave polarized in the $z$--direction is
\begin{equation}\label{redshiftEMwavezpolar}
z_z\approx \frac{1}{a(t_e)}\left[1+\eta_z(t_e)\right]-1\, .
\end{equation}

Redshifts \eqref{redshiftEMwavexpolar}, \eqref{redshiftEMwaveypolar} and \eqref{redshiftEMwavezpolar} are all different, as $\eta_x\neq\eta_y\neq\eta_z\neq \eta_x$, in general. This occurs because
each polarization couples differently to the anisotropic spacetimes. Also, the redshifts are now dispersive as they depend non-trivially on the wavelenghts of the EM waves.
This is not surprising as EM waves do not propagate along geodesics, and therefore, waves propagating in different directions behave differently.

This effect is not present if the cosmology is isotropic, as when $\epsilon=0$, then $\eta_x=0=\eta_y=\eta_z$ by Eqs.~\eqref{solutionetaX}, \eqref{solutionetaY} and \eqref{solutionetaZ}.  In such cases, the isotropic FRW--like light propagation is in null geodesics, and its corresponding redshift, are recovered \cite{misner,wald}.

\section{Discussion on wavelength-dependent redshifts}

If the Universe is isotropic, described by a FRW metric, the cosmological redshift of light does neither depend on the direction of incoming  light, on its wavelength nor on its polarization. However, if the Universe is anisotropic, the previous statement is no longer valid. As we showed in previous sections, in any anisotropic case, the redshift of light depends on two different features: the direction of propagation of light with respect to the principal axes of spacetime, and the polarization of the EM wave.

In Sec.~\ref{sectionnull}, where light is considered as an EM wave under the geometrical optics approximation (light ray), it has been shown that different directions of propagation of light yield different redshifts. Even more, if light propagates in a general direction, not only along a principal axis,  the redshift becomes dispersive (wavelength dependent) for rays arriving from different directions.

Even in a more general fashion, if the EM wave is studied beyond the geometrical optics limit, the polarization of the wave plays an essential role. In Sec.~\ref{sectionnonnull}, it was shown that when EM wave propagation is studied by solving the complete Maxwell equations.  Thereby, the resultant redshifts depend on the direction of propagation and on the polarization with respect to that direction, and therefore, they also may be dispersive, depending  on the wavelength of EM wave  and the relative direction   between  the  propagation vector of the EM wave and the anisotropic axes.

All the previous effects have their origin in the spacetime anisotropy, and in that way, they can be used as a tool to detect any possible cosmological anisotropy of the Universe in its early stages.
Several researches have focused in determining, in an indirect manner, the effects of anisotropy on redshifts \cite{ferreras,nodl,menezes}.
In general, those observations indicate that our Universe is almost isotropic, as limited by current experimental capabilities. This implies that the anisotropy is, if any, very small. However, as it was discussed in previous section, any anisotropy, no matter how small, introduces new effects on redshifts.

Recently, it has been measured the wavelength dependence of the cosmological redshift \cite{ferreras}. Any possible dependence will introduce a  correction $\Delta z$ in the such way the redshift will acquire the form
\begin{equation}
\frac{\omega(t_e)}{\omega(t_o)}=\left[1+z_{\tiny{\mbox{FRW}}}\right]\left[1+\Delta z(t_e)\right]\, ,
\end{equation}
where $z_{\tiny{\mbox{FRW}}}=a(t_o)/a(t_e)-1$, is the FRW cosmological redshift. In Ref.~\cite{ferreras}, it has been measured that the $\Delta z\sim 10^{-6}$, or below,  with the statistical uncertainty of their procedure. This is an indication that our Universe is almost isotropic.

According to our results of previous sections,
if light is considered in the high--frequency limit (geometrical optics approximation) in  the small anisotropic cosmological model, then from \eqref{redshiG} we can see that
\begin{equation}\label{redshiGDeltaz}
\Delta z(t_e)=-\frac{K_y^2}{K^2}\epsilon(t_e)\, ,
\end{equation}
is a direct consequence of the anisotropy of the Universe.  Thereby, different  directions of the light propagation will induce different $\Delta z$. For  light propagating in a direction perpendicular to the anisotropy axis, $\Delta z=0$. But for light propagation parallel to the anisotropy axis, then $\Delta z=-\epsilon$. This imposes a constraint on the anisotropy of spacetime. If results of Ref.~\cite{ferreras} are considered, then  we can infer that $|\epsilon|\leq 10^{-6}$. For a general direction of propagation of a light ray, then a measurement should give $\Delta z\ll  10^{-6}$.

More generally, if light is considered as a EM wave, then for a EM wave with $j$--polarization (with $j=x,y,z$), then
\begin{equation}\label{redshiGDeltazj}
\Delta z_j(t_e)=\eta_j(t_e)\, ,
\end{equation}
where $\eta_j$ can be given by Eqs.~\eqref{solutionetaX}, \eqref{solutionetaY} or \eqref{solutionetaZ}, depending on the polarization as well as of the wavelength. Again, by Ref.~\cite{ferreras}, we infer that $|\eta_j|\leq 10^{-6}$, but expecting to measure different redshift corrections $\Delta z_j$ for different polarizations.

Although, the redshift corrections \eqref{redshiGDeltaz} and \eqref{redshiGDeltazj} are not equal due to the nature of the solutions to Maxwell equations, they both predict that in an anisotropic universe redshift depends on the direction of propagation and it is dispersive. Light with different wavevectors and wavelengths interacts with the anisotropy of the spacetime.
If any experiment detects both these effects, then any evidence of small cosmological anisotropy can be established.

In addition to these, redshift correction \eqref{redshiGDeltazj} contains also information on the polarization of light and the structure of the spacetime. This redshift is valid for large wavelength electromagnetic waves. Thus, if an experiment is focused on detecting such EM waves, any possible correction on the FRW redshift can also give a hint on the global and local temporal dynamics of the anisotropy by Eqs.~\eqref{solutionetaX}, \eqref{solutionetaY} and \eqref{solutionetaZ}. Through the derivatives of $\epsilon$ in $\eta$,  the (local or cosmological) scale lengths of the anisotropy can also be determined.
On the other hand, a possible measurement of spacetime anisotropy can be based on the comparison of the redshift for two different polarizations. The quantity $\Delta z_i(t_e)-\Delta z_j(t_e)=\eta_i(t_e)-\eta_j(t_e)$ should be nonzero for large wavelength EM waves in any anisotropic spacetime. In this way, several possible experiments can be used to study the level of isotropy of our Universe.

Finally, the results \eqref{solutionetaX}, \eqref{solutionetaY}, \eqref{solutionetaZ} and \eqref{redshiGDeltazj} for EM waves, establish that the plane of polarization can rotate. This can be deduced from the amplitudes \eqref{amplitudEMwavex}, \eqref{amplitudEMwavey} and \eqref{amplitudEMwavez}, which oscillates on time through the form of $\eta_j$ for each polarization. This effect coincides with the controversial observational results first noticed in Ref.~\cite{nodl}. This phenomenon will be theoretically studied in a forthcoming article dealing with EM wave solutions to Maxwell equations.


\begin{thebibliography}{}

\bibitem{misner} C. W. Misner, K. S. Thorne and J. A. Wheeler, {\it Gravitation} (W. H. Freeman and Co. San Francisco, 1973).
\bibitem{wald} R. M. Wald, {\it General Relativity} (The University of Chicago Press, 1984).
\bibitem{ryan} M. P. Ryan, Jr. and L. C. Shepley, {\it  Homogeneous Relativistic Cosmologies} (Princeton University Press, Princeton, 1975).
\bibitem{hojman1} F. A. Asenjo and S. A. Hojman, Phys. Rev. D {\bf 96}, 044021 (2017).
\bibitem{hojman2} F. A. Asenjo and S. A. Hojman, Class. Quantum Grav. {\bf 34}, 205011  (2017).
\bibitem{carroll} S. Carroll, {\it Spacetime and Geometry, An introduction to General Relativity} (Addison Wesley, San Francisco, 2004).
\bibitem{oron} {{\O}}. Gr{{\o}}n and S. Hervik, {\it Einstein’s General Theory of
Relativity: With Modern Applications in Cosmology} (Springer, New York, 2007).
\bibitem{ryden} B. Ryden, {\it Introduction to Cosmology} (Addison Wesley, 2003).

\bibitem{ferreras} I. Ferreras and I. Trujillo, ApJ {\bf 825}, 115 (2016).
\bibitem{nodl} B. Nodland and J. P. Ralston, Phys. Rev. Lett. {\bf 78}, 3043 (1997).
\bibitem{menezes} R. S. Menezes Jr. and C. Pigozzo and S. Carneiro, JCAP  {\bf 2013}, 033 (2013).

\end{thebibliography}
\end{document}